\documentclass[twocolumn,secnumarabic,amssymb, nobibnotes, aps, prd]{revtex4}
\usepackage{graphicx}

\begin{document}
\title{Observation of dc power on system of identical asymmetric superconducting rings.}
\author{V. L. Gurtovoi$^{1,2}$} \author{V.N. Antonov$^{2,3}$} \author{M. Exarchos$^{3}$} \author{A. I. Il'in$^{1}$} 
\author{A.V. Nikulov$^{1}$}
\affiliation{$^{1}$Institute of Microelectronics Technology and High Purity Materials, Russian Academy of Sciences, 142432 Chernogolovka, Moscow District, RUSSIA.} 
\affiliation{$^{2}$Moscow Institute of Physics and Technology, 29 Institutskiy per., 141700 Dolgoprudny, Moscow Region, Russia}
\affiliation{$^{3}$Physics Department, Royal Holloway University of London, Egham, Surrey TW20 0EX, UK}
\begin{abstract} The observations the dc voltage on asymmetric superconducting ring testify that one of the ring segments is a dc power source. The persistent current flows against the total electric field in this segment. This paradoxical phenomenon is observed when the ring or its segments are switched between superconducting and normal state by non-equilibrium noises. We demonstrate that the dc voltage and the power increase with the number of the identical rings connected in series. Large voltage and power sufficient for practical application can be obtained in a system with a sufficiently large number of the rings. We point to the possibility of using such a system for the observation of the dc voltage above superconducting transition and in the asymmetric rings made of normal metal. 
 \end{abstract}

\maketitle

\narrowtext

\section{Introduction}
\label{}
An electric current induced in a resistive circuit $R > 0$ will rapidly decay in the absence of an applied voltage. But the persistent current $I_{p}$, quantum phenomenon observed in rings of superconductors \cite{LP1962,Science2007} and normal metals \cite{Science2009PC,PRL2009PC} may stay alive for long time under these conditions. It is known that the conventional circular electric current $I$ induces the potential difference
$$V = 0.5(R _{n}-R _{w}) I  \eqno{(1)}$$                                                        
on the halves of the ring with different resistance $R _{n} > R _{w}$. Could the persistent current induce a similar voltage? The observations \cite{LP1962,Science2007,Science2009PC,PRL2009PC} $I_{p} \neq 0$ at $R > 0$ allow to answer on this question experimentally. The voltage $V_{dc} \propto I_{p}$ can easily be distinguished since the persistent current changes periodically in magnetic field $B$ with the period $B _{0} = \Phi _{0}/S$ corresponding to the flux quantum $\Phi _{0} = 2\pi \hbar /q$ inside the ring with the area $S$ \cite{LP1962,Science2007,Science2009PC,PRL2009PC}. The flux quantum equals $\Phi _{0} \approx 20.7  \ Oe \ \mu m^{2}$ for Cooper pairs with the charge $q = 2e$. The quantum oscillations of the dc voltage $V_{dc}(\Phi ) \propto I_{p}(\Phi )$ were observed first time as far back as 1967 in measurements of an asymmetric dc SQUID, i.e. a superconducting loop with two Josephson junctions \cite{Physica1967}, and later in measurements of an asymmetric superconducting ring \cite{NANO2002}. This effect still remains without the due attention. Although the observation $V_{dc}(\Phi ) \propto I_{p}(\Phi )$ deserves the attention due to its paradoxicality and possible practical significance. The experiments \cite{Physica1967,NANO2002} testify that the persistent current $I_{p}$, circulating in the ring clockwise or anticlockwise, flows against the dc voltage $V_{dc}$, directed from left to right or from right to left, in one of the ring halves. Therefore this half is a dc power source $V_{dc}I_{p}$. In order to draw the attention of experimenters to the importance of the quantum effect discovered in \cite{NANO2002,Physica1967} we demonstrate that the dc power $V_{dc}I_{p}$ is summed in a system of identical asymmetric superconducting rings connected in series. The amplitude of the oscillations $V_{dc}(\Phi )$, which was observed on the asymmetric dc SQUID, was not exceeded $15 \ \mu V$  \cite{Physica1967} and $1 \ \mu V$ at the measurement of the single asymmetric ring \cite{NANO2002}. The amplitude can be increased many times when using a system with a large number of identical rings.   

\begin{figure}
\includegraphics{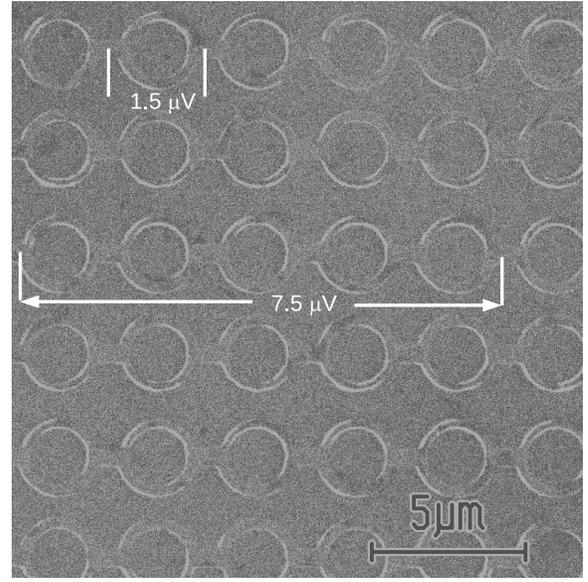}
\caption{\label{fig:epsart} A fragment of the system of 1080 identical asymmetric aluminum rings with $r \approx 1 \ \mu m $ connected in series. The semi-ring widths are $w _{w} \approx 400 \ nm$ and  $w _{n} \approx 200 \ nm$. The voltage $V\approx 7.5 \ \mu V$ is observed of the five rings connected in series when the voltage $V\approx 1.5 \ \mu V$ is observed on each ring.}
\end{figure} 

\section{Experimental Details}
 \label{}
 We use the system of 1080 asymmetric aluminum rings with the same radius $r \approx 1 \ \mu m $, Fig.1,  and the system of 667 aluminum rings with the same radius $r \approx 0.5 \ \mu m $. The systems were fabricated by e-beam lithography and lift-off process of $d \approx 20 \ nm$ (the 1080 rings) and $d \approx 30 \ nm$ (the 667 rings) thick aluminum film. The 1080 rings were more asymmetric than the 667 one: the arm widths of all 1080 rings were $w _{n} \approx 200 \ nm$ and $w _{w} \approx 400 \ nm$ for narrow and wide parts, respectively and $w _{n} \approx 100 \ nm$ and $w _{w} \approx 125 \ nm$ of all 667 rings. The resistance in the normal state was $R _{n} \approx  8000 \ \Omega$ of the 1080 rings and $R _{n} \approx  5400 \ \Omega$ of of the 667 rings; the resistance ratio $R(300 K)/R(4.2 K) \approx  1.7$ and $R(300 K)/R(4.2 K) \approx  2$; superconducting transition temperature $T _{c} \approx 1.360 \ K$ and $T _{c} \approx 1.320 \ K$; the width of the resistive transition $\Delta T_ {c} \approx 0.02 \ K$ and $\Delta T_ {c} \approx 0.01 \ K$. The temperature dependence of the critical current is described by the relation $I _{c} = I _{c}(T=0)(1 - T/T _{c})^{3/2}$, Fig.4, where $I _{c}(T=0) \approx 700 \ \mu A$ for the 1080 rings $I _{c}(T=0) \approx 520 \ \mu A$ for the 667 rings. The critical current density $j _{c}(T=0) \approx 10^{7} A/cm^{2}$ equals approximately the depairing current density \cite{PCJETP07}.

\section{A transformer without the primary winding}
 \label{}
The oscillations $V_{dc}(\Phi ) \propto I_{p}(\Phi )$ with the amplitude up to $V _{A} \approx 1 \ \mu V = 10^{-6} \ V$ were observed on a single asymmetric aluminum ring at the temperature $T \approx 0.98T _{c}$ and non-equilibrium noises $I _{noise} = \int _{fmin}^{fmax} df I _{f} \sin (2\pi ft + \phi _{f})$ with the current amplitude $\overline{2I _{noise}^{2}}^{1/2} = (\int _{fmin}^{fmax} df I _{f}^{2})^{1/2} \approx 3 \ \mu A$. We observed the oscillations, Fig.2, with the amplitude $V _{A} \approx 1.6 \ mV = 0.0016 \ V$ about a thousand times more on the 1080 rings under similar conditions. The increase of the voltage $V _{N} = NV _{1}$ with the number $N$ of the rings is a trivial effect if the current $I$ in the ring with the resistance $R _{n} + R _{w}$ is induced by the Faraday electromotive force $-d\Phi /dt $ in accordance with the well known relation $(R _{n} + R _{w})I = -d\Phi /dt $. A homogeneous magnetic field $B$ induces the same magnetic flux $\Phi  = BS = \pi r^{2}$ in all rings with the same radius $r$. Therefore the current $I = (R _{n} + R _{w})^{-1}(-d\Phi /dt)  = (R _{n} + R _{w})^{-1}S(-dB/dt)$ has the same direction and value in all identical rings when the magnetic field changes in time $dB/dt  \neq 0$. The voltage (1) should also has the same sign and value in all identical rings and therefore $V _{N} = NV _{1}$. For example, if a current $I$ induces the voltage (1) $V_{1} \approx 1.5 \ \mu V$ on each ring then the voltage  $V_{5} \approx 7.5 \ \mu V$ will be observed on five rings connected in series, Fig.1, and the voltage  $V_{1080} \approx 1620 \ \mu V = 1.62 \ mV$ should be observed on the 1080 rings. 

\begin{figure}
\includegraphics{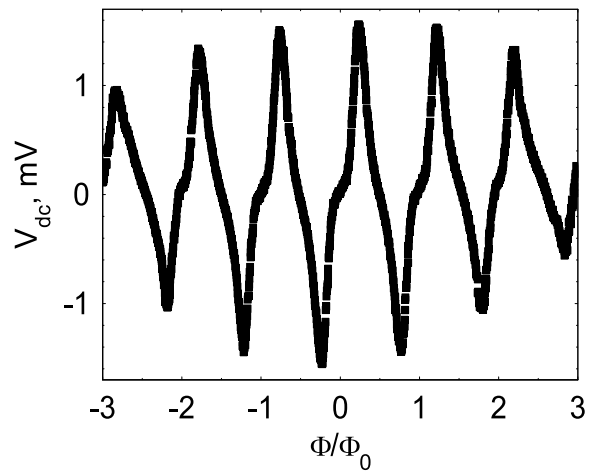}
\caption{\label{fig:epsart} Quantum oscillations of the dc voltage in magnetic field observed on the system of 1080 aluminium rings at the temperature $T \approx 1.330 \ K \approx 0.978T _{c}$ when the rings are switched between superconducting and normal states by the ac current with the amplitude $I _{A} \approx 3 \ \mu A$. The period $B _{0} = \Phi_{0}/S \approx 5.2 \ Oe$ corresponds to the area $S = \pi r^{2}  \approx 20.7/5.2  \approx 4 \ \mu m^{2}$ of the rings with $r \approx 1.1 \ \mu m$ shown on Fig.1.}
\end{figure} 

The system of asymmetric rings may be considered in this case as the secondary winding of the electric transformer. The primary winding of the electric transformer induces the current $I _{sw}$ in the secondary winding in order to obtain the power $W _{load} = R _{load}I_{sw}^{2}$ on a load with the resistance $R _{load}$, Fig.3 at the left. The circular current $I _{sw}$ flows in the secondary winding against the potential electric field $E _{p} = -\nabla V$ thanks to the Faraday electromotive force $-N_{sw}d\Phi /dt $ induced by the current $I _{pw}$ in the primary winding, Fig.3 at the left. Here $N_{sw}$ is the number of coils of the secondary winding; $\Phi $ is the magnetic flux induced by the current $I _{pw}$ of the primary winding. The total electric field $E = -\nabla V - dA/dt$ equals zero in the secondary winding when its resistance is equal zero. The voltage on the load $V _{load} = R _{load}I_{sw}$ equals the Faraday electromotive force $V _{load} = -N_{sw}d\Phi /dt $ in this ideal case. 

\begin{figure}
\includegraphics{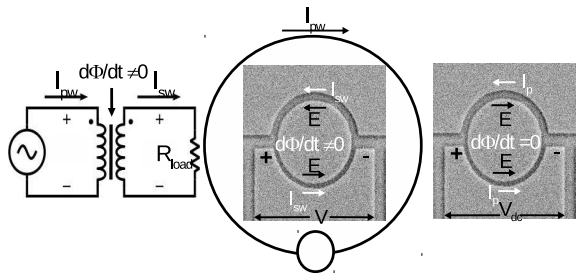}
\caption{\label{fig:epsart} The transformer diagram is shown on the left. The current $I _{pw}$ of the primary winding creates magnetic flux varying in time $d\Phi /dt $ in order to induce the Faraday electromotive force and the current $I _{sw}$ in the secondary winding and the current $I _{sw}$ and the voltage $V _{load} = R _{load}I_{sw}$ in the load with the resistance $R _{load}$. The current $I _{pw}$ of the primary winding can induce the Faraday electromotive force and the current $I _{sw}$ also in the asymmetric ring shown in the center. In this case the wide half may be considered as the secondary winding and the narrow half as the load. On the right is the case of the dc voltage observed on an asymmetric ring with the persistent current at a magnetic flux not varying in time $d\Phi /dt = 0$. The persistent current $I _{p}$ flows against the total electric field $E = -\nabla V$ in the wide half in contrast to the conventional current $I _{sw}$ the direction of which corresponds to the direction of the total electric field $E = -\nabla V - dA/dt$ thanks to the Faraday electric field $-dA/dt = -l^{-1}d\Phi /dt$, see in the center. The photos of a real aluminum ring with the radius $r \approx 1 \ \mu m$ is shown. Such ring was used for the observation of the $V _{dc}(\Phi )$ oscillations.}
\end{figure}

The circular current $I _{sw}$ flows against the potential electric field $E _{p} = -\nabla V$ also in the wide half when the current $I _{pw}$ of the primary winding induces the Faraday electromotive force in the asymmetric ring, Fig.3 at the center. Therefore the wide half with the resistance $R _{w}$ may be considered as the secondary winding whereas the narrow half with the resistance $R _{n} > R _{w}$ may be considered as the load. Useful power can be obtained at a load connected in parallel to the narrow half, although such a transformer is not the most efficient. The voltage and power obtained on the load must increase in proportion to the number of asymmetric rings connected in series. The system of asymmetric rings connected in series cannot be a power source without the primary winding when the conventional circular current in the ring is induced by the Faraday electromotive force. The persistent current is observed without the Faraday electromotive force, Fig.3 at the right. Therefore the system of asymmetric rings with the persistent current can be a power source without the primary winding.

\section{The persistent current flows against the total electric field}
 \label{}
Our observation is not trivial and is even paradoxical since we observed the dc voltage $V _{dc} \approx 1.6 \ mV$ in magnetic field, for example $B \approx  \Phi _{0}/4S$, constant in time $dB/dt = 0$, Fig.2. The conventional electric current, induced by the Faraday electromotive force $-d\Phi /dt = -ldA/dt$, flows against the potential electric field $E = -\nabla V$ in the wide half (1), but its direction corresponds to the direction of the total electric field $E = -\nabla V - dA/dt$ in both halves, Fig.3 at the center. The Ohm law $j\rho = E = -\nabla V - dA/dt$ and the forces balance are valid in this case: $F _{E} + F _{dis} = 0$ at $dI/dt = 0$. The force $F _{E} = qE$ of the electric field $E = -\nabla V - dA/dt$ acting on the electrons is balanced the dissipation force $F _{dis}$. These laws are not valid for our observation, Fig.2. The observation of the dc voltage  $V _{dc} $ at $-d\Phi /dt = -ldA/dt = 0$ means that the persistent current flows against the total electric field $E = -\nabla V $ in the wide half, Fig.3 at the right.

\section{Why can the persistent current not decay?}
 \label{}
This paradox is connected with other paradox: the persistent current $I _{p} \neq 0$ does not decay in the ring with a non-zero resistance $R > 0$ \cite{LP1962,Science2007,Science2009PC,PRL2009PC} in spite of the energy dissipation with the power $RI _{p}^{2}$ without the Faraday electromotive force $-d\Phi /dt = 0$. The persistent current of Cooper pairs is not zero $I _{p} \neq 0$ but the resistance is equal zero $R = 0$ in superconducting state, whereas in normal state $R > 0$ but $I _{p} = 0$. Therefore the both paradoxes $V_{dc}(\Phi ) \propto I_{p}(\Phi )$, Fig.2, and $I_{p} \neq 0$ at $R > 0$ \cite{LP1962,Science2007} can be observed when the ring or its segments are switched between superconducting and normal states. 

The wave function $\Psi = |\Psi |e^{i\varphi } $ describing a superconducting pairs exists along the whole circle $l = 2\pi r$ when all ring's segments are in superconducting state with a non-zero density of Cooper pairs $|\Psi |^{2} = n _{s} > 0$. According to the canonical definition, the gradient operator $\hat{p} = -i\hbar \nabla $  corresponds to the canonical momentum $p = mv + qA$ of a particle with a mass $m$ and a charge $q$ both with $A \neq 0$ and without $A = 0$ magnetic field. Whereas the operator of the velocity $\hat{v} = (\hat{p} - qA)/m = (-i\hbar \nabla - qA)/m$ \cite{LandauL} depends on the magnetic vector potential $A$. The angular momentum of each Cooper pair in the superconducting state has discrete values $m _{p} =  \oint_{l}dl  \Psi ^{*}\hat{p}\Psi /2\pi \oint_{l}dl  \Psi ^{*}(-i\hbar \nabla )\Psi /2\pi = \hbar n$ due to the Bohr quantization $m_{p} = rp = rmv = n\hbar$ or the requirement $\oint _{l} dl \nabla \varphi = n2\pi $ of uniqueness of the wave function at any point of the circle $\Psi = |\Psi |e^{i\varphi } =  |\Psi |e^{i(\varphi + n2\pi )} $. The velocity 
$$\oint_{l}dl v = \oint_{l}dl \frac{\hbar \nabla \varphi  - qA}{m}  = \frac{\hbar}{mr} (n - \frac{\Phi }{\Phi_{0}})\eqno{(2)}$$
cannot be equal zero when the magnetic flux $\Phi = \oint_{l}dl A$ inside the ring is not divisible $\Phi \neq n\Phi _{0}$ by the flux quantum $\Phi _{0}$ due to the dependence of  the operator of the velocity on the magnetic vector potential $A$. The effects connected with this dependence were first predicted by Aharonov and Bohm \cite{AB1959}. Therefore, they are referred as the Aharonov - Bohm effects. 

The persistent current equals  
$$I_{p} = \frac{q\hbar}{mr\overline{(s n _{s})^{-1}}} (n - \frac{\Phi }{\Phi _{0}}) = \frac{\Phi_{0}}{L_{k}}(n - \frac{\Phi }{\Phi _{0}})  \eqno{(3)}$$ 
since all $N _{s} = \oint _{l}dl s n _{s}$ Cooper pairs, being bosons, have the same quantum number $n$ in superconducting ring with the macroscopic volume  $V = \oint _{l}dl s $. Here $\overline{(s n _{s})^{-1}} = l^{-1}\oint _{l}dl (s n _{s})^{-1}$ and $L_{k} = ml\overline{(s n _{s})^{-1}}/q^{2}$ is the kinetic inductance of Cooper pairs in the ring with the section area $s$ and the density of Cooper pairs $n _{s}$ varying along the circumference $l = 2\pi r$. The value of the persistent current (3) and the discreteness of the permitted state spectrum depend on the density of Cooper pairs $n _{s}$ in each ring's  segment \cite{PLA2012QF}. The difference the kinetic energy \cite{Tink75}
$$E_{n} = \frac{L_{k}I_{p}^{2}}{2} =  \frac{\Phi_{0}^{2}}{2L_{k}}(n - \frac{\Phi }{\Phi_{0}}) ^{2} \eqno{(4)}$$ 
between the permitted states is large $E_{n+1} - E_{n} \approx \Phi_{0}^{2}/2L_{k} \gg k _{B}T$ when $n _{s} > 0$ \cite{NanoLet2017} in all ring's segments. But it becomes zero when a segment $l _{A}$ is switched in normal state with $n _{s,A} = 0$ and $R _{A} > 0$ since $1/L_{k} \propto 1/\overline{(s n _{s})^{-1}} \rightarrow 0$ at $n _{s,A} \rightarrow 0$ \cite{PLA2012QF}. The current, circulating in the ring $I(t) = I _{p}\exp -t/\tau _{RL}$ should decay during a short relaxation time $\tau _{RL} = L/R _{A}$ after this transition at $t = 0$, here $L$ is the total inductance of the ring. But the persistent current (3) must appear again due to the quantization (2) when all ring's segments return in superconducting state since the state with the zero current is forbidden at $\Phi \neq n'\Phi_{0}$. The current will have the same direction at $\Phi \neq (n'+0.5)\Phi_{0}$ because of the predominate probability of the permitted state $n$ corresponding to the minimal kinetic energy (4). Therefore the current $\overline{I_{p}} =  \Theta ^{-1}\int _{\Theta }dtI(t) \neq 0$ average in time $\Theta  \gg 1/f _{sw}$ is observed at a non-zero resistance average in time  $\overline{R_{l}} =  \Theta ^{-1}\int _{\Theta }dtR_{l}(t) > 0$ \cite{LP1962,Science2007} when the ring or its segments are switched between superconducting and normal states with a frequency $f _{sw}$. According to the theoretical prediction and the experimental results \cite{Science2007} this current is diamagnetic at $n'\Phi_{0} < \Phi < (n'+0.5)\Phi_{0}$, paramagnetic at $(n'+0.5) \Phi_{0} < \Phi < (n'+1)\Phi_{0}$ and equal zero at $\Phi = n'\Phi_{0}$ and $\Phi = (n'+0.5)\Phi_{0}$. The current $\overline{I_{p}}$ changes periodically in magnetic field $B$ with the period $B _{0} = \Phi_{0}/S$ \cite{LP1962,Science2007} due to the change of the quantum number $n$ corresponding to the minimal kinetic energy (4).

\section{The DC voltage induced by switching a ring's segment between superconducting and normal states.}
 \label{}
The voltage (1) should be equal zero in a symmetric ring with the equal resistance of the halves $R _{n} = R _{w}$. The voltage should not be observed also in a symmetric ring with the persistent current when its segments are switched in normal state with the equal frequency. The potential difference $V_{A}(t) = R _{A}I(t) = R _{A}I _{p}\exp -(t - t _{i})/\tau _{RL}$ should appear on a segment $l _{A}$ after each its transition at $t = t _{i}$ in the normal state with the resistance $R _{A}$. The sign of the voltage $V_{A}$ will correspond to the direction of the persistent current $I _{p}$ each time $t _{i}$. But the dc voltage will not be observed if all other ring's segments are switched also as the segment $l _{A}$. The dc voltage $V _{dc} = \overline{V} =  \Theta ^{-1}\int _{\Theta }dtV(t) \neq 0$ can be observed only in an asymmetric ring with dissimilar segments. For example, the dc voltage $V _{dc}  \approx f _{sw}L\overline{I_{p}}$ should be observed at the low frequency of the switching $f _{sw} \ll 1/\tau _{RL}$ and $V _{dc}  \approx R _{A}\overline{I_{p}}$ at the high frequency $f _{sw} \gg 1/\tau _{RL}$ when only one segment $l _{A}$ is switched in normal state \cite{LTP1998}.

The persistent current $\overline{I_{p}} \neq 0$ is observed at a non-zero resistance $\overline{R_{l}} > 0$ \cite{LP1962,Science2007} in the temperature region corresponding to the superconducting resistive transition where $0 < \overline{R_{l}} < R _{n}$ because of thermal fluctuations switching ring's segments between the superconducting state (with $R = 0$) and normal state (with $R = R _{n}$) \cite{Tink75}. The oscillations of the dc voltage $V_{dc}(\Phi ) \propto \overline{I_{p}}(\Phi )$ were observed \cite{Physica1967,NANO2002} in superconducting state at $T < T _{c}$ where thermal fluctuations cannot switch ring's segments in normal state. They are switched in this case by non-equilibrium noises. The experimental investigation \cite{PCJETP07,Letter2003} have corroborate that the oscillations $V_{dc}(\Phi )$ appear when the amplitude of the noises or a sinusoidal current $I _{sin} = I _{A} \sin (2\pi f t)$ reaches the critical current $I _{c} = I _{c}(T=0)(1 - T/T_{c})^{3/2}$ at the temperature $T  < T _{c}$ of measurement. Their amplitude quickly reaches a maximum and decreases with further increase in the current amplitude \cite{PCJETP07,Letter2003}. The temperature dependence of the $V_{dc}(\Phi )$ amplitude $V_{A}(T)$ is also non-monotonic for a given value of the amplitude $\overline{2I _{noise}^{2}}^{1/2}$ \cite{Letter2007,PL2012PC,APL2016}. The oscillations $V_{dc}(\Phi )$ appear when the critical current $I _{c}(T)$ decreases down to $\overline{2I _{noise}^{2}}^{1/2}$, the amplitude $V_{A}(T)$ increases with temperature $T$, reaches a maximum $V_{A,max}$ at $T = T_{max}$, and then decreases \cite{Letter2007,PL2012PC,APL2016}. 

The maximum voltage $V_{A,max}$ increases and is observed at a lower temperature $T_{max}$ with the increase of the amplitude $\overline{2I _{noise}^{2}}^{1/2}$ because of the temperature dependence of both the critical current and the persistent current, see Fig.4. The maximum power $V_{dc}I _{p} \propto (1 - T/T_{c})^{2}$ increases with the temperature decrease when $\overline{2I _{noise}^{2}}^{1/2} > I _{c}$ because $I _{p,A} \propto 1 - T/T _{c}$ and $V_{A,max} \propto I _{p,A}$. We observed the dc voltage $V_{dc} \approx 1.5 \ \mu V$ on each ring and $V_{dc} \approx 1.6 \ mV$ on 1080 rings at $B \approx \Phi_{0}/4S$, Fig.2, and at $T \approx  0.98T _{c}$ when the persistent current $I _{p} \approx I _{p,A}/2 \approx 0.6 \ \mu A$, Fig.4. This values corresponds  to the power $V_{dc}I _{p} \approx 2 \ 10^{-12} \ W$ on each ring and $V_{dc}I _{p} \approx 2 \ 10^{-9} \ W = 2 \ nW $ on 1080 rings. This power is observed when the rings is switched between the superconducting and normal states by the noise with the amplitude $\overline{2I _{noise}^{2}}^{1/2} \approx 3 \ \mu A$. The power should increase with the $\overline{2I _{noise}^{2}}^{1/2}$ increase. 

\section{Detector of weak noise and dc power source }
 \label{}
The non-equilibrium noise $\overline{2I _{noise}^{2}}^{1/2} \approx 3 \ \mu A$ switching the single ring in the normal state at $T \approx 1.2 \ K < T _{c}$ in \cite{NANO2002} is the thermal Nyquist noise equilibrium at the room temperature $T \approx 300 \ K $. The power of the Nyquist noise $W _{Nyk,f} = 4k _{B}Tdf$ is distributed evenly across all frequencies from zero  $f _{min} = 0$ to the quantum limit $f _{max} \approx k _{B}T/h$ \cite{Feynman}. The total power of the Nyquist noise $W _{Nyk,t} = 4(k _{B}T)^{2}/h$ at the room temperature $T \approx 300 \ K $ reaches $ \approx 10^{-7} \ W$. Here $k _{B} \approx 1.4 \ 10^{-23} \ J/K$ is the Boltzmann constant, $h \approx 6.6 \ 10^{-34} \ J s$ is the Planck constant. The current amplitude $\overline{2I _{noise}^{2}}^{1/2} \approx 3 \ \mu A$ corresponds to the total power of the Nyquist noise at an effective resistance $R _{hf}  \approx W _{Nyk,t}/\overline{2I _{noise}^{2}}$ of the wires connecting the room's and low-temperature measuring system equal $\approx 10 \ k \Omega$. The noise amplitude was reduced by an order of magnitude down to $\overline{2I _{noise}^{2}}^{1/2} \approx 0.25 \ \mu A$ in \cite{Letter2007}, due to the increase in the effective resistance at high frequencies up to $R _{hf}  \approx 1 \ M \Omega$. The effective resistance was increased up to $R _{hf}  \approx 100 \ M \Omega$ and the noise amplitude was decreased down to a value of less than $\overline{2I _{noise}^{2}}^{1/2} \approx 20 \ nA$ in \cite{PL2012PC} thanks to special low-temperature $\pi$-filters and coaxial resistive twisted pairs.

\begin{figure}
\includegraphics{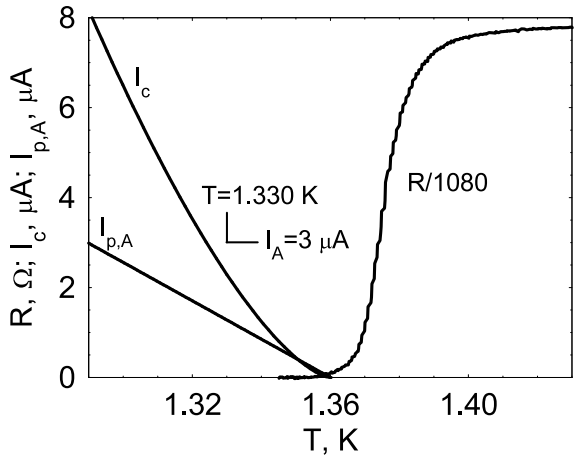}
\caption{\label{fig:epsart} Temperature dependence of the resistance $R$ (superconducting resistive transition),  of the critical current $I _{c} = I _{c}(T=0)(1 - T/T_{c})^{3/2} $ and the amplitude $I _{p,A} = I _{p,A}(T=0)(1 - T/T_{c})$ of the system of 1080  aluminium rings. $I _{c}(T=0)  \approx 700 \ \mu A$, $I _{p,A}(T=0) \approx 58 \ \mu A$, $T _{c} \approx 1.36 \ K$. The temperature $T \approx  1.330 \ K$ and the current amplitude $I _{A} = 3 \ \mu A$ at which the dc voltage oscillations are observed, Fig.2, are indicated.}
\end{figure}

This noise reduction allowed to demonstrate the possibility of using asymmetric superconducting rings connected in series as a detector of very weak noise \cite{APL2016}. Here we draw the attention of experimenters on the opportunity to use the system with big number of asymmetric superconducting nano - rings as the dc power source. This exploitation requires the increase rather than the reduction of the noise amplitude $\overline{2I _{noise}^{2}}^{1/2}$. The load at the room temperature $T \approx 300 \ K $ may be considered also as the source of the Nyquist current in the electric circuit including the system of asymmetric superconducting rings. The Nyquist current switches the rings between superconducting and normal states and thus induces the dc voltage $V_{dc}$.

\section{The ratio between the critical current and the persistent current}
 \label{}
The maximum power $V_{dc}I _{p} \propto I _{p}^{2}$ is observed when $\overline{2I _{noise}^{2}}^{1/2} > I _{c}(T,B)$. Therefore one can get more power $V_{dc}I _{p}$ at low noise $\overline{2I _{noise}^{2}}^{1/2}$ when the critical current is much less than the persistent current $I _{c}(T,B) \ll I _{p}$. According to the predictions of the theory, confirmed experimentally \cite{JETP07J}, the critical current of the symmetric ring is described by the formula  
$$I _{c} = I _{c0} - 2| I _{p}| = I _{c0} - 2 I _{p,A}2|n - \frac{\Phi}{\Phi _{0}}| \eqno{(5)}$$
The ratio $I _{p,A}/I _{c0} =  \surd 3\xi (T)/4r$ of the critical currents $I _{c0} = I _{c0}(T=0)(1 - T/T_{c})^{3/2}$ at $I _{p} = 0$ to the amplitude $I _{p,A} = I _{p,A}(T=0)(1 - T/T_{c})$ the persistent current is determined by the ratio of the correlation length of the superconductor $\xi (T) = \xi (0)(1 - T/T _{c})^{-1/2}$ to the radius of the ring $r$ \cite{Tink75}. Therefore the critical current at $\Phi \approx (n + 0.5)\Phi _{0}$ of the ring with a small radius $r \approx \surd 3\xi (T)/2$ may be equal zero or be much smaller than the persistent current $I _{c} \approx I _{c0} - 2I _{p,A} \ll I _{p,A}$.  Measurements of the system of 667 of aluminum rings with the radius $r \approx 500 \ nm$, Fig.5, corroborate this possibility. The magnetic dependence of the critical current measured in the opposite directions are almost identical, Fig.5, because the rings are almost symmetric $w _{w} \approx w _{n}$. The dependence are described by the relation (6) at $I _{c0} \approx  2.9 \ \mu A$ and $I _{p,A} \approx  1.25 \ \mu A$. The relation $I _{p,A}/I _{c0} =  \surd 3\xi (T)/4r \approx 0.43$ corresponds to $\xi (T=0.97T _{c}) \approx 500 \ nm$ and the value $\xi (0) \approx 100 \ nm$ typical for aluminium film with small free path of electrons. The theoretical $I _{c} \approx I _{c0} - 2I _{p,A} \approx 0.4 \ \mu A$ and measured $I _{c} \approx 0.6 \ \mu A$ at $\Phi = \pm 0.5\Phi _{0}$ values of the critical current is smaller the persistent current $|I _{p}| = I _{p,A} \approx  1.25 \ \mu A$.

\begin{figure}
\includegraphics{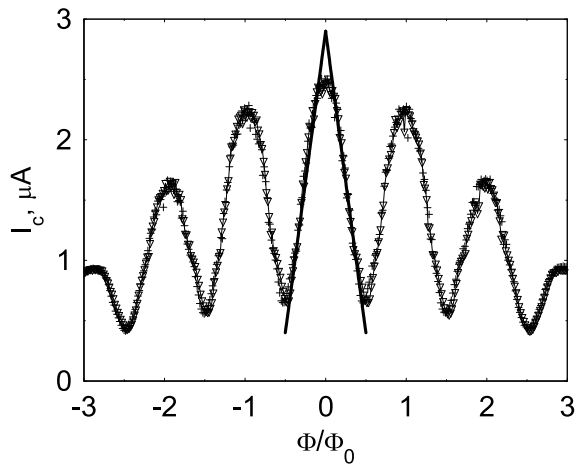}
\caption{\label{fig:epsart} Quantum oscillations in magnetic field of the critical current of 667 rings measured at the temperature $T \approx  1.283 \ K \approx 0.972T _{c}$ in the opposite directions $I _{c+}(\Phi /\Phi _{0})$ (triangles) and $I _{c-}(\Phi /\Phi _{0})$ (crosses). The experimental dependence correspond to the theoretical one for the symmetric rings (5) with $I _{c0} \approx  2.9 \ \mu A$ and $I _{p,A} \approx  1.25 \ \mu A$ (lines). The discrepancy between theoretical and experimental values near  $\Phi = 0$ may be explained the influence of the contacts between rings the width of which is smaller than the total width of two ring-halves. The period $B _{0} = \Phi_{0}/S \approx 22.6 \ Oe$ corresponds to the area $S = \pi r^{2}  \approx 20.7/22.6 \approx 0.9 \ \mu m^{2}$ of the rings with $r \approx 0.54 \ \mu m$ used for the measurements.}
\end{figure}

\section{Quantum force}
The connection of the observations $\overline{I_{p}} \neq 0$ at $\overline{R_{l}} > 0$ \cite{LP1962,Science2007} and $V_{dc}(\Phi ) \propto \overline{I_{p}}(\Phi )$, Fig.2,   with the switching between the discrete and continuous spectrum of the permitted states (4) allows to describe why the persistent current does not decay in spite of a non-zero dissipation $\overline{R_{l}I_{p}^{2}} > 0$ and can flow against the total electric field  $E = -\nabla V_{dc} $. The angular momentum of each from $N _{s}$ pairs changes from the value $m _{p} = \hbar n$ corresponding to the quantization to the value $m _{p} = q \Phi /2\pi = \hbar \Phi /\Phi _{0}$ corresponding to the zero velocity $v = 0$ when the circular electric current $I(t)$ changes from $I(t) = I _{p}$ to $I(t) = 0$. This change occurs under the influence of the dissipation force $F _{dis}$. The opposite change from the $m _{p} = \hbar \Phi /\Phi _{0}$ to $m _{p} = \hbar n$ should occur due to the quantization when the entire ring returns in the superconducting state. The change
$$F_{q} = \hbar  (\overline{n} - \Phi /\Phi _{0})f _{sw}/r \eqno{(6)}$$ 
of the momentum $p$ per an unit time due to the quantization when the ring is switched with a frequency $f _{sw}$ is called "quantum force" in \cite{PRB2001}. The quantum force compensates for the dissipation force and provides a balance of forces 
$$2\pi rF _{q} + \oint _{l}dl F _{dis} = 0 \eqno{(7)}$$ 
replacing the Faraday electromotive force $-qd\Phi /dt + \oint _{l}dl F _{dis} = 0$.  

\section{Could the dc voltage be observed in the fluctuation region of superconducting asymmetric rings and in normal metal asymmetric rings?}
According to (1) the voltage should observed when the current $I$ in the asymmetric ring and its resistance $R _{n} > R _{w}$ are not zero. The persistent current $\overline{I_{p}} \neq 0$ is observed at a non-zero resistance $\overline{R_{l}} > 0$ in the fluctuation region of superconductors rings at $T \approx T _{c}$ \cite{LP1962,Science2007} and in normal metal rings \cite{Science2009PC,PRL2009PC}. Whereas the voltage $V_{dc}(\Phi ) \propto \overline{I_{p}}(\Phi )$ was observed \cite{Physica1967,NANO2002} for the present in the main only in superconducting state where the equilibrium resistance $R_{l} = 0$.  It is more difficult to observe the voltage $V_{dc}(\Phi ) \propto \overline{I_{p}}(\Phi )$ at $\overline{R_{l}} > 0$ since the persistent current is much smaller in the fluctuation region and in normal metal rings than in superconducting state: in the superconducting state $I _{p,A} \approx 100 \ \mu A (1- T/T _{c})$ \cite{PCJETP07} whereas in the fluctuation region $I _{p,A} \approx 0.1 \ \mu A$ \cite{Science2007,Letter2007} and $I _{p,A} \approx 0.001 \ \mu A$ \cite{Science2009PC} in normal metal rings. Therefore the 1080 rings were needed in order to observe the oscillations $V_{dc}(\Phi ) $ with the amplitude $V _{A} > 0.02 \ \mu V$ in the lower part of the resistive transition $\overline{R_{l}} < 0.3R _{n}$ \cite{PL2012PC}.   

It is needed more rings in order to observe the visible oscillations $V_{dc}(\Phi ) $ in the upper part of the resistive transition and in normal metal rings. The experimental investigations of the system with big number of asymmetric rings connected in series may have fundamental importance. The authors \cite{Science2009PC} note fairly: "{\it An electrical current induced in a resistive circuit will rapidly decay in the absence of an applied voltage. This decay reflects the tendency of the circuit's electrons to dissipate energy and relax to their ground state}" and claim that the persistent current is dissipationless in spite of the non-zero resistance of the rings. The author \cite{Birge2009} agrees with the authors \cite{Science2009PC} although he recognizes: "{\it The idea that a normal, nonsuperconducting metal ring can sustain a persistent current - one that flows forever without dissipating energy - seems preposterous. Metal wires have an electrical resistance, and currents passing through resistors dissipate energy}".

The authors \cite{Science2009PC} claim that the dissipation power  equals zero $R _{l}I _{p}^{2} = 0$ although they measure a non-zero resistance  $R _{l} > 0$ and observe the persistent current $I _{p} \neq 0$. They don't even try to explain the contradiction of their claim, according to which $R _{l}I _{p}^{2} = 0$ at $R _{l} > 0$ and $I _{p} \neq 0$, with mathematics. The opinion of the author \cite{Kulik1970n} (who has predicted in the first time the persistent current in normal metal) about the paradoxical possibility $I _{p} \neq 0$ at $R _{l} > 0$ does not contradict mathematics: "{\it The current state corresponds in this case to the minimum of free energy, so the account of dissipation does not lead to its disintegration}". It is argued in \cite{Kulik75} that the author \cite{Kulik1970n} rather than the authors \cite{Science2009PC,Birge2009} is right. According to his opinion the $I _{p} \neq 0$ observed at $R _{l} > 0$ is a type of the Brownian motion \cite{Feynman} likewise the Nyquist noise. Nobody claims that the Brownian motion and its type - the Nyquist current are dissipationless. The kinetic energy of Brownian's particles dissipates into the thermal energy $k _{B}T$ and is taken from the thermal energy \cite{Feynman}. Therefore the power of the Nyquist noise $W _{Nyk} = 4k _{B}T\Delta f = R _{l}I _{N,f}^{2}$ is proportional to the thermal energy \cite{Feynman}. The authors \cite{Science2009PC,Birge2009} claim that the power of the persistent current equals zero $W _{p} = R _{l}I _{p}^{2} = 0$ since it is the power of the direct current in contrast to the power of the Nyquist noise which equals zero at the zero frequency $f = 0$. According to their claim the voltage (1) should not be observed in spite of a non-zero value $R _{n} - R _{w}$ and the persistent current $I _{p}$ in normal metal ring \cite{Science2009PC,PRL2009PC} since the observation of the voltage $V$ means the observation of the power $VI _{p}$. In contrary to the opinion of the authors \cite{Science2009PC,PRL2009PC} the voltage $V \propto I _{p}$ may be observed according to the author \cite{Kulik75}. 

\section{Conclusion}
The authors \cite{PRL1990} used a system of approximately ten million ($N \approx  10000000$!) of rings with the radius $r \approx 300 \ nm$ in order to observe the persistent current of electrons. The proportionality $V _{A,N} \approx NV _{A,1} $ of the voltage $V _{A,N}$ with the number $N$ of asymmetric rings connected in series means that the oscillation $V_{dc}(\Phi )$ with the amplitude up to $10 \ V$ may be observed on the system with such huge number of rings when the rings are switched between superconducting and normal states by the noises with the amplitude $\overline{2I _{noise}^{2}}^{1/2} \approx 3 \ \mu A$. The $10^{7}$ rings occupy a $7 \ mm^{2}$ area on the substrate \cite{PRL1990}. The obvious relation (1) raises the question: "Can the persistent current observed above the superconducting transition and in normal metal rings create a potential difference in asymmetric rings?" The system with a large number of identical asymmetric rings should allow to answer on this question experimentally. The answer will have fundamental importance. It is more difficult to observe the voltage in the case of the rings made of normal metals since the persistent current of electrons, in contrast to the one of Cooper pairs, has different direction even in identical rings.

\section*{Acknowledgement} This work was made in the framework of State Task No 007-00220-18-00 and has been supported by the Russian Science Foundation, Grant No. 16-12-00070.

\end{document}